# Methods to Estimate Advanced Driver Assistance System Penetration Rates in the United States

Noah Goodall, Ph.D., P.E.

Virginia Transportation Research Council, USA, noah.goodall@vdot.virginia.gov
ORCID: 0000-0002-3576-9886

August 1, 2024

*Preprint version, under review.*

---

**Abstract**

Advanced driver assistance systems (ADAS) are increasingly prevalent in the vehicle fleet, significantly impacting safety and capacity. Transportation agencies struggle to plan for these effects as ADAS availability is not tracked in vehicle registration databases. This paper examines methods to leverage existing public reports and databases to estimate the proportion of vehicles equipped with or utilizing Levels 1 and 2 ADAS technologies in the United States. Findings indicate that in 2022, between 8% and 25% of vehicles were equipped with various ADAS features, though actual usage rates were lower due to driver deactivation. The study proposes strategies to enhance estimates, including analyzing crash data, expanding event data recorder capabilities, conducting naturalistic driving studies, and collaborating with manufacturers to determine installation rates.

---

# 1   Introduction

Advanced Driver Assistance Systems (ADAS) have become increasingly prevalent on roadways. These systems, which include features such as adaptive cruise control, lane keeping assist, and automatic emergency braking, rely on radar, lidar, and machine vision to assist drivers in vehicle operation.

ADAS technologies have been shown to significantly impact roadway safety and capacity. Kim et al. (*1*) demonstrated through microscopic simulation that a freeway segment with 100% adaptive cruise control usage could increase capacity by 28% compared to conventional vehicles, while a 40% penetration rate yielded only a 10% increase. Cicchino (*2*) found that forward collision warning combined with automated emergency braking reduced rear-end striking crashes by 50%. Similarly, lane departure warning was associated with a 21% reduction in injury crashes (*3*).

ADAS's impact on traffic and safety is dependent on the market penetration of the various technologies, yet transportation agencies cannot easily estimate the prevalence of these systems. Government vehicle registration systems do not include fields for ADAS technologies. Although all new vehicles are assigned a Vehicle Identification Number (VIN) indicating the make, model, year, and place of manufacture, nothing in the current VIN standard represents ADAS functions. Even sales figures can be misleading, as many features are sold as optional packages, and studies have shown that even when ADAS features are equipped, drivers may elect to use them at highly variable rates. Without accurate estimates of ADAS penetration, governments cannot adequately plan for the anticipated impacts of these technologies.

To address the challenge, this study has three objectives:

1. Identify ADAS features of greatest significance for transportation agencies based on their effects on safety and operations.
2. Identify methods to estimate installed and usage rates of ADAS systems using available data.
3. Propose methods to estimate rates using alternative or non-public data.

The scope of this study is limited to the United States, but findings may be applicable to other countries with similar vehicle fleets, databases, and VIN formats. The results of this study will benefit transportation agencies and researchers by providing more accurate estimates on the rate of ADAS in the vehicle fleet, allowing the development and calibration of existing empirical models of the effect of these technologies on traffic flow, capacity, safety, and infrastructure planning. As ADAS technologies continue to evolve and proliferate, understanding their penetration rates becomes increasingly crucial for effective transportation planning and policy-making.

# 1   Literature Review

## 1.1   ADAS Taxonomy

Before recording or estimating ADAS penetration rates, a consistent terminology for the range of ADAS technologies must first be established. Various classification schemas for ADAS exist in the literature.

The most widely-used classification scheme is the SAE levels of driving automation (*4*). First introduced in 2014 and most recently revised in 2021, the SAE guidance classifies vehicle automation technologies into six levels. Vehicles operating at Levels 1–2 have one or more advanced driver assistance systems (ADAS). ADAS is distinguished from higher levels of automation by the requirement for a human driver to be responsible for object and event detection and response.

Gasser et al. (*5*) proposed a different classification scheme for ADAS, based on the functionality of the system. Their three broad groups are:

- Operation type A: informing warnings and functions, e.g., lane departure warning
- Operation type B: continuously automating functions, e.g., adaptive cruise control
- Operation type C: intervening emergency functions (near-crash situations), e.g., automatic emergency braking

In a study funded by the U.S. Department of Transportation, Gouribhatla and Pulugurtha (*6*) conducted a literature review of the effect of ADAS on driver behavior. As part of their study, they identified eight unique ADAS technologies: blind spot warning, lane departure warning, over speed warning, forward collision warning, adaptive cruise control, cooperative ACC, lane keeping assist, and automated emergency braking. Similarly, Pradhan et al. (*7*) reviewed press releases and vehicle user manuals from 30 manufacturers and identified 207 systems that could be classified as one of 51 ADAS features.

## 1.2   Rates of ADAS Technologies Installed

Estimating the rate of ADAS technologies installed in vehicles presents significant challenges. The United States does not have any laws requiring manufacturers to report ADAS installations in production vehicles (*8*), so features added to production vehicles are often unknown to government agencies in the absence of voluntary reporting by the manufacturer.

The Highway Loss Data Institute (HLDI), a sister organization of the Insurance Institute for Highway Safety (IIHS), has provided estimates of various vehicle safety features beginning in 2012 (*9*), with annual updates between 2014 and 2023. HLDI maintains a database of feature availability by vehicle make, model, and year, which is then compared against their database of over 490 million unique VINs to determine the proportion of vehicles in the U.S. fleet with features that are either standard or optional (*10*). The rate of installation for optional features is estimated "based on known take rates from some manufacturers" (*11*). In a letter to the National Highway Traffic Safety Administration (NHTSA), IIHS indicated that these take rates were based on "special samples" voluntarily provided to IIHS by "some manufacturers" (*12*). HLDI further extrapolates future adoption rates of these technologies based on historical adoption rates and vehicle registration data from IHS Markit, now S&P Global (*13*). The IHS Markit data is not public but is available for purchase. HLDI estimated that in 2022, approximately 28% of the United States vehicle fleet was equipped with front crash prevention warning, 24% with lane departure warning, 22% with automatic emergency braking, and 4% with combined adaptive cruise control and lane centering (*11*). Estimates for adaptive cruise control are not provided separately, but only as part of a combined adaptive cruise control and lane centering systems.

Pradhan et al. (*7*) attempted to determine ADAS penetration rates using a database of VINs of vehicles registered in Massachusetts but were unsuccessful. The authors described the difficulties they encountered, noting that there is no straightforward and direct method to access information about the ADAS features in various models even after discussions with insurance providers and the Alliance of Automobile Manufacturers. They instead proposed randomly sampling vehicle registration data and cross referencing any available manufacturer data, a process they described as "painstaking and burdensome" (*7*).

Gajera et al. (*14*) analyzed fatal crashes from 2016 to 2019 using NHTSA's Fatality Analysis Reporting System (FARS). The study examined 138,899 vehicles involved in fatal crashes, decoding their VINs through the NHTSA Product Information Catalog and Vehicle Listing (vPIC) database to identify vehicles with optional or standard lane keeping assistance, lane centering, or adaptive cruise control features. Their findings indicated that 2,428 vehicles (1.8%) had at least one of these technologies available as either an optional or standard feature.

The study's results, however, are limited by NHTSA's data collection methods for automation technologies. While 49 CFR Part 565 does not mandate manufacturers to report the prevalence of automation technologies, some do so voluntarily. NHTSA independently researches automation features from press releases and vehicle manuals, but this effort is restricted to vehicles from model year 2017 onwards, produced by 38 major light-duty vehicle manufacturers (*15*).

These limitations significantly impact the Gajera et al. (*14*) analysis. Data for 2016 and early 2017 model year vehicles would be severely limited, as few vehicles with recorded vPIC automation data would have been in operation. The analysis would also exclude many pre-2017 vehicles equipped with automation features, as these would not be captured in the vPIC data.

## 1.3  Rates of ADAS Technology Usage

While vehicles may have ADAS technologies installed, individual drivers may choose to disable passive features or not engage active features while driving. To estimate the percentage of vehicles equipped with ADAS technology that have not deactivated the technology, Reagan et al. (*16*) observed vehicles brought in for service at 14 Washington, D.C. metro area dealerships. The researchers found activation rates of 93% for vehicles equipped with automatic emergency braking, 57% for vehicles equipped with adaptive cruise control, and 8% for vehicles equipped with lane departure prevention.

Flannagan et al. (*17*) collected data from approximately 2,000 General Motors vehicles equipped with forward collision warning and lane departure warning driven 19 million miles on public roads. Participants were observed to have activated lane departure warning approximately 50% of driving time, and forward collision warning 91% of driving time.

From driver surveys, the main factors that affect usage rates of ADAS appear to be driver annoyance and understanding of system capabilities (*18*). Specifically, forward collision prevention warnings that are well-understood by drivers and that have low false alarm rates are used at higher rates, while lane departure prevention systems with less annoying alerts and consistent pavement marking detection are used more frequently. Other studies suggest that vibratory warnings were seen as less annoying than auditory warnings (*16*, *17*).

## 1.4  ADAS Registration Systems

Based on analysis of state vehicle automation laws (*8*)no states were identified that record vehicle automation features directly as part of vehicle registration. This lack of systematic recording of ADAS features in vehicle registration systems presents a significant challenge in accurately estimating ADAS penetration rates across the United States.

Several states have begun to include automation fields on police crash reports. Pennsylvania's crash reporting form has a field for automation status, with an option of "partial automation" defined as " driver assist functions available at the time of the crash such as blind spot detection, lane departure warning, adaptive cruise control, collision avoidance braking, etc." (*19*). Texas also records vehicle automation levels using SAE's definitions, but the reporting so far suggests potential underreporting. Only 0.23% (n = 4645) of vehicles involved in crashes between March 2023 and July 2024 were reported to have any automation present (*20*), significantly below HLDI estimates (*11*). When automation was present, it was Level 1 in 41% of vehicles, Level 2 in 11%, and unknown in 48%, suggesting there may be some challenges in collected comprehensive data at the roadside. As it stands, the data quality of data from state police crash records, even when the data is available, will not support ADAS penetration rate estimates in the near future.

## 2    Estimating ADAS Penetration

Estimating market penetration of ADAS technologies presents significant challenges due to inconsistent nomenclatures, varying operational scopes, and the absence of these technologies in standard vehicle registration databases. For most vehicles, determining which ADAS technologies are installed without direct inspection is not possible. This section presents two methods to estimate ADAS prevalence using existing data sources.

### 2.1    Selection of Most Relevant ADAS Features

The ADAS features of most significance to transportation agencies are those with the greatest impact on agency missions of supporting safe and efficient transportation on public roads. Research indicates that warning systems have lesser impacts on crash rates than continuous automation systems and crash prevention systems. Data from on-road studies found only a 16% reduction in rear-end striking crashes for forward collision alerts, but a 45% reduction for front automatic braking (*21*). Similarly, lane departure warnings resulted in a 3% reduction in lane departure crashes, but lane keep assist resulted in a 30% reduction (*21*). Other studies have reported similar findings (*2*). Data on continuous automation and crash prevention systems should therefore be prioritized over warnings systems.

Level 2 ADAS involves the combination of lateral and longitudinal vehicle control, which can be represented as a combination of two separate ADAS systems, i.e., lane centering assist and adaptive cruise control. By tracking individual ADAS systems, a database can record both Level 1 and Level 2 vehicles, as Level 2 vehicles are simply vehicles that allow simultaneous use of these lateral and longitudinal control ADAS functions. Adaptive cruise control has also been shown through simulation to have significant effects on freeway capacity at high usage rates (*1*). Adaptive cruise control and lane centering can be considered high priority features for tracking.

Forward collision prevention systems such as automatic emergency braking have been shown to reduce rear-end striking crashes by 45% (*21*), representing some of the highest safety improvements of any ADAS technology. For this reason, forward collision prevention systems can also be considered high priority.

Using these guidelines, the highest value ADAS features for transportation agencies are adaptive cruise control, automatic emergency braking, forward collision prevention, lane centering assist, lane departure prevention, and pedestrian automatic emergency braking. The remainder of the study focuses on these technologies.

## 2.2 HLDI Estimated Equipped Rates

The Highway Loss Data Institute (HLDI) (*11*) has developed a method to estimate penetration rates for various ADAS features that overcomes the limitations of standard VIN encoding. HLDI obtains special samples of VINs from vehicle manufacturers for vehicles equipped with specific ADAS technologies, particularly for models where these features are optional.

These manufacturer-provided VINs are cross-referenced with vehicle registration data maintained by IHS Markit, now S&P Global (*13*), to estimate the proportion of vehicles equipped with each technology. For optional features, HLDI calculates the probability of installation using regression models that consider factors such as model year, vehicle size, class, and base price. This approach allows for more accurate estimates of actual ADAS penetration rates, particularly for optional features that may have varying take rates across different vehicle models. It also enables the evaluation of ADAS effectiveness by linking the VIN-level technology information to crash and insurance claims data. HLDI's (*11*) most recent estimates for front crash prevention, lane departure warning, automatic emergency braking, and combined adaptive cruise control and lane centering are shown in Figure 1.

In their reports, HLDI predicts the prevalence of penetration rates in future years, out to 2050. To evaluate the accuracy of these predictions, predictions of 2022 vehicle equipped rates from the 2018 report (*22*) were compared with the 2022 estimates in the 2023 report (*11*). The results are shown in Figure 1. In all cases, predictions were highly accurate, even when technologies increased by 400% in the vehicle fleet over the five-year timeline.

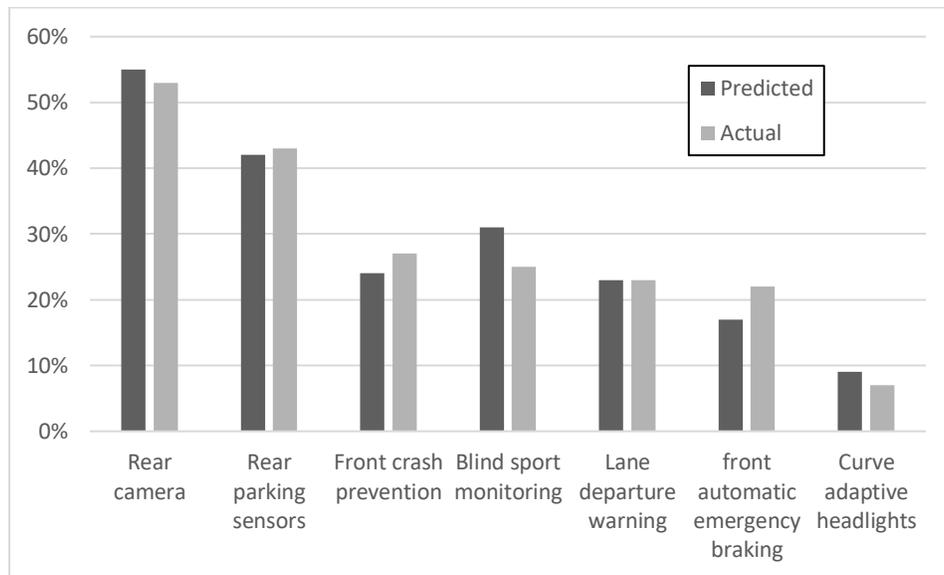

**Figure 1. HLDI Predictions of 2022 Vehicle Equipped Rates from 2018 vs. HLDI Estimated 2022 Equipped Rates from 2022.**

*2.3    Estimating from Comparable HLDI New Vehicle Installation Rates of Similar Technologies*

Adaptive cruise control, lane keep assist, lane centering, and pedestrian automatic emergency braking rates were not estimated in the HLDI report. Two of these features, adaptive cruise control and lane keep assist, are discussed in a different HLDI (*10*) report showing the prevalence of these technologies in new model year vehicles. This report, however, does not estimate prevalence of the feature in the entire vehicle fleet, but rather in new model year vehicles only.

Fortunately, the new vehicle report (*10*) also includes new vehicle equipped rates of technologies discussed in the ADAS prevalence reports (*11*). By identifying technologies with similar new vehicle adoption rates as adaptive cruise control and lane keep assist, the total fleet equipped rate of adaptive cruise control and lane keep assist can be estimated as a function of the equipped rate of the matching technology.

For instance, HLDI (*10*) reported that 15% of 2020 model year vehicles had adaptive cruise control as a standard feature, with another 55% offering it as an option. The same report showed that 16% of 2018 model year vehicles had lane departure warning as standard, with 55% offering it optionally. These rate comparisons are shown in Figure 2. Assuming similar market entry rates for these technologies, the 2020 fleet equipped rate for lane departure warning approximates the 2022 rate for adaptive cruise control. Given that the 2020 fleet equipped rate for lane departure warning was 16% (*11*), we can estimate the 2022 equipped rate for adaptive cruise control at 16% as well. The adjustment is shown graphically in Figure 3.

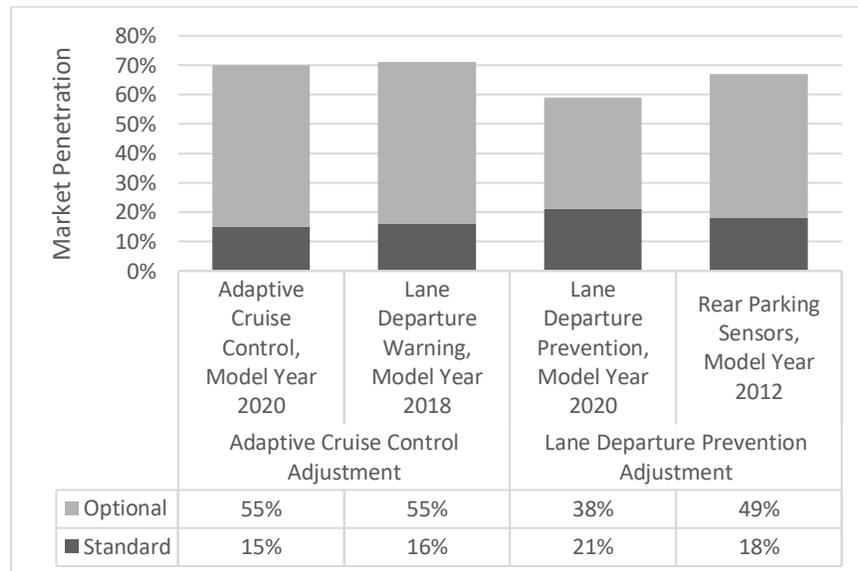

**Figure 2. Model year standard and optional rates for different ADAS features, recreated from (*10*).**

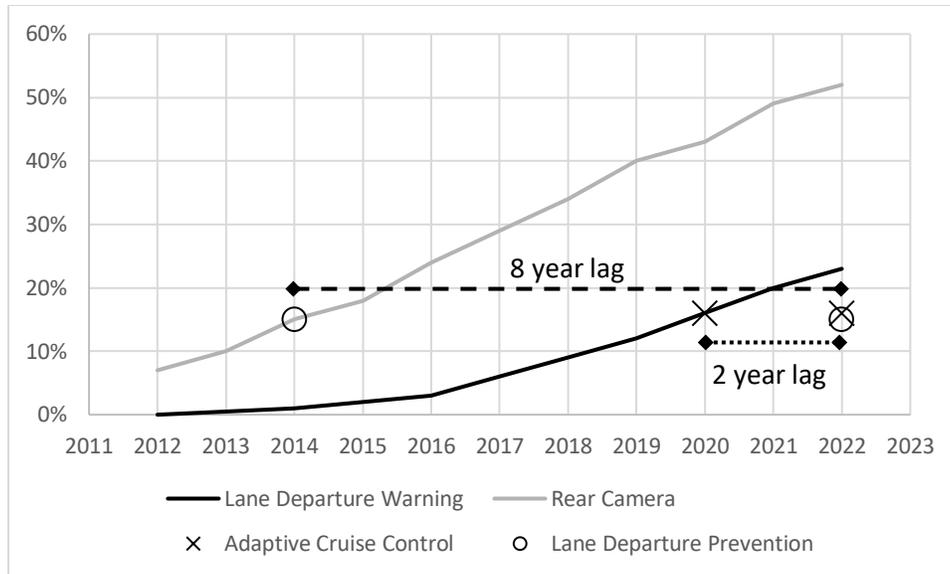

**Figure 3. Adjustment of ADAS features based on install rates of similar features in prior years, based on data from (*11*).**

A similar method estimates lane departure prevention system rates. In 2020 models, 21% had these systems as standard, with 38% offering them optionally (*10*). This pattern resembles 2012 model year rear parking sensors, which were standard on 18% and optional on 49% of vehicles. These rate comparisons are shown in Figure 2. Assuming an eight-year lag in adoption rates, the 2014 fleet equipped rate for rear parking sensors of 15% (*11*) can serve as an estimate for the 2022 rate of lane departure prevention systems.

There are limitations with using this methodology. This approach assumes that technologies are adopted at similar rates, which may not occur in reality. Equipped rates of most technologies increase by 2–4 percentage points per year (*11*), and small differences in adoption rates can be substantial when compounded over a long time period. Additionally, some technologies are more likely to be installed as features. Front automatic emergency braking, for example, has a much higher percentage of optional installations than rear parking sensors (*11*). Finally, some technologies may be subject to government mandates or voluntary agreements which may accelerate adoption compared to technologies not subject to mandates. One example is electronic stability control, which was subject to a mandate first announced in 2007 and effective 2012. Caution should be used when extrapolating equipped rates of technologies over long periods of time, when comparing against technologies with different rates of optional inclusion, and technologies subject to mandates or voluntary agreements.

### 2.4 Estimating from FARS-vPIC New Vehicle Installation Rates of Similar Technologies

Not all of the ADAS features of interest are included in the HLDI estimates. For example, lane centering assist and pedestrian automatic emergency braking were not included in the HLDI (*10*) model year reports, and so estimating the equipped rates of these features requires additional calculation. The equipped rate by model year of these other features can instead be estimated by analyzing the automation features of vehicles involved in fatal crashes in the FARS dataset. FARS provides VINs for each involved vehicle, which can be decoded in vPIC, which reports whether a feature was sold as standard or optional on that particular model year. The percentage of model 2021 vehicles equipped with lane centering assist and pedestrian automatic emergency braking involved in fatal 2021 crashes are shown in Figure 4.

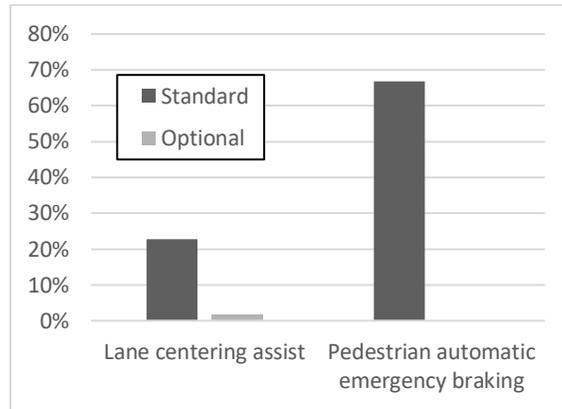

**Figure 4. Percentage of model year 2021 vehicles involved in 2021 fatal crashes with ADAS features.**

Analysis of FARS-vPIC data reveals that 23% of model year 2021 vehicles involved in 2021 fatal crashes had lane centering assist as a standard feature, with an additional 2% offering it as an option. HLDI (*10*) data shows that in 2003, 18% of vehicles had electronic stability control as standard, with 13% offering it optionally. Assuming lane centering assist follows a similar market entry pattern to electronic stability control, and that fatal crashes represent the broader vehicle fleet, we can estimate that the 2022 fleet equipped rate for lane centering assist mirrors the 2004 rate for electronic stability control. Based on this comparison, the 2022 equipped rate for lane centering assist is estimated at 8% (*23*).

A comparable approach estimates pedestrian automatic emergency braking rates. FARS-vPIC data indicates that 67% of 2021 model year vehicles in fatal crashes had this feature as standard, with no optional offerings. HLDI (*10*) reports that in 2008, 61% of vehicles had electronic stability control as standard, with 14% offering it optionally. If pedestrian automatic emergency braking adoption aligns with electronic stability control, and assuming fatal crashes reflect the overall fleet composition, the 2022 fleet equipped rate for pedestrian automatic emergency braking likely corresponds to the 2009 rate for electronic stability control. Given that electronic stability control's 2009 fleet equipped rate was 25% (*23*), we can estimate the 2022 rate for pedestrian automatic emergency braking at 25% as well. These estimates are reported in the first column of Table 1.

This approach assumes that installation rates over time were similar for different technologies. Lane centering assist and pedestrian automatic emergency braking have never been mandated nor subject to a voluntary agreement among automakers. Electronic stability control, however, became mandatory on new vehicles beginning in 2012 based on Federal Motor Vehicle Safety Standard (FMVSS) 126 (*24*) which was announced in 2006 (*25*). The calculations for lane centering assist relies on electronic stability control equipped rates from 2004, and were likely unaffected by FMVSS 126. The pedestrian automatic emergency braking estimates relies on electronic stability control equipped rates from 2009, which may have been higher than expected due to the FMVSS 126 mandate.

*2.5    Activation Rate of Installed Features*

To estimate the percentage of drivers that have the capability to use ADAS features but leave them deactivated, Reagan et al. (*16*) observed vehicles brought in for service at 14 Washington, D.C. metro area dealerships. The researchers found activation rates of 93% for vehicles equipped with automatic emergency braking, 57% for vehicles equipped with adaptive cruise control, and 8% for vehicles equipped with lane departure prevention. For features not included in their study, lane centering assist can be assumed to match adaptive cruise control and pedestrian automatic emergency braking to match automatic emergency braking due to the similarities between paired features.

The Reagan et al. (*16*) study was conducted in 2016 and may not reflect current activation rates. Technologies may have become more user-friendly in recent years, leading to higher activation rates. Their study was also conducted in the Washington D.C. metro region, and results may not apply to rural regions. Their findings agree with a General Motors study from conducted between 2013 and 2014 (*17*) which found a 91% activation rate for forward collision warning, compared to Reagan et al.'s (*16*) 93% two years later.

*2.6    Features Installed and Activated*

Market penetration estimates for ADAS technologies across all registered vehicles in the United States in 2022 are presented in Table 1. The data is expressed in three categories:

- Vehicles equipped: the percentage of all vehicles estimated to have the technology installed.
- Activated when equipped: the percentage of equipped vehicles where the technology is not deactivated, based on observations at 14 dealership service centers in the Washington, DC metro region (*16*).
- Activated of total fleet: The product of the equipped rate and activation rate, representing the estimated percentage of the total vehicle fleet with the technology both installed and not deactivated.

It should be noted that the activated fleet percentage represents an upper bound of potential usage. Actual utilization may be considerably lower, as drivers with activated systems (such as adaptive cruise control) may still choose not to engage them in many driving scenarios.

**Table 1. Estimated 2022 Market Penetration of Vehicle Automation Technologies in United States**

| Technology | Vehicles Equipped | Activated when Equipped | Activated of Total Fleet |
|---|---|---|---|
| Adaptive cruise control | 16%‡ | 57%† | 9% |
| Automatic emergency braking | 16%* | 93%† | 15% |
| Forward collision prevention | 22%* | 93%† | 20% |
| Lane centering assist | 8%§ | 57%** | 5% |
| Lane departure prevention | 15%‡ | 65%† | 10% |
| Pedestrian automatic emergency braking | 25%§ | 93%** | 23% |

* From HLDI (*11*) equipped vehicle estimates
† From observed activation rates at 14 dealership service centers (*16*)
‡ Comparing leading adoption rates of similar technologies from model year rates of target technology (*10*)
§ Comparing leading adoption rates of similar technologies from model year rates of target technology involved in fatal crashes (*10*, *15*, *26*)
** Assumed same as similar technologies in Reagan et al. (*16*)

# 3   PROPOSED APPROACHES

Given the challenges in estimating ADAS penetration rates using current methods, several approaches are proposed to improve data collection and estimation accuracy. These approaches leverage existing data sources, potential future datasets, and alternative methods of data collection.

## 3.1   VIN Decoding of Registered Vehicles

One approach involves decoding VINs of all registered vehicles, or a representative sample, using NHTSA's vPIC database. This method can yield information on standard or optional ADAS features for specific vehicle makes, models, and years.

The vPIC database primarily comprises manufacturer-reported data, submitted in compliance with 49 CFR Part 565. NHTSA enhances this data through independent research for certain areas of interest (*15*). This includes scrutinizing manufacturer websites, press releases, and vehicle manuals to identify ADAS features that may be standard or optional equipment across various vehicle specifications.

This approach is relatively straightforward and efficient, requiring only a few weeks to process VIN queries using the vPIC application programming interface (API). A key limitation is that vPIC's ADAS feature data only extends back to 2017 model years. For earlier model years, HLDI estimates could serve as a proxy, offering a reasonable approximation of market penetration.

A significant challenge with this method is determining the actual installation rate of optional technologies, which remains unknown and would require estimation based on HLDI (*11*) rates for comparable technologies. Despite these constraints, this approach could establish a baseline understanding of ADAS penetration rates. It also offers the advantage of being maintainable as a standalone database, subject to annual updates.

## 3.2 Leveraging HLDI Estimates

The Highway Loss Data Institute produces the most sophisticated estimates of ADAS penetration rates currently available. Transportation agencies could directly apply these annually published estimates. This approach requires minimal effort, as HLDI compiles and releases these figures.

The national-level presentation of HLDI data, rather than state-specific breakdowns, presents a potential limitation. The impact of this limitation is likely minimal, as individual state vehicle fleets are not expected to deviate significantly from the national fleet composition beyond the error margins inherent in the existing assumptions.

HLDI estimates do not cover all ADAS features of interest, notably excluding adaptive cruise control, lane keep assist, and lane centering. A separate HLDI report (*10*) does provide installation rates for adaptive cruise control and lane keep assist in new vehicles. These data can be utilized to estimate fleet penetration rates by comparing them with new vehicle installation rates of analogous technologies for which total fleet penetration rates are known.

## 3.3 VIN Reports by Manufacturer

Estimates could potentially be obtained directly from manufacturers. This approach would involve sending the VINs of all or a sample of registered vehicles to manufacturers, who could then provide information on whether certain technologies are installed. This method would eliminate the need to estimate the proportion of vehicles with optional features that are actually installed.

This approach has the advantage of potentially providing vehicle-specific installation rates. If the data is comprehensive, it may provide the highest quality estimates. However, there are significant challenges to this approach. It is unclear which manufacturers have these data, are willing to participate, and can provide accurate estimates. The fact that NHTSA must manually collect these data from press releases and vehicle manuals suggests that manufacturers may be unwilling to share vehicle-specific ADAS installation rates with government agencies.

This approach also requires significant effort as there are numerous manufacturers to coordinate with and potentially significant data cleaning required. However, if successful, this method could provide the most accurate and comprehensive ADAS penetration data.

## 3.4 Estimates of Actual On-Road Usage

Even though vehicles may have automation features installed, drivers may deactivate systems or leave them activated but rarely used in driving. Activation rates can be estimated from studies in the literature, of which the current best source is Reagan et al. (*16*). Researchers may be able to estimate actual on-road usage rates from future naturalistic driving studies or similar studies.

### 3.4.1 Roadside Detection of ADAS Sensor Signals

Many ADAS technologies use sensors to detect nearby vehicles, vulnerable road users, and obstacles. Theoretically, roadside sensors could detect lidar, radar, and sonar signals from passing vehicles to determine ADAS equipment and usage. Several factors render this approach impractical. First, transmissions from several systems running simultaneously create noise, making determination of whether a vehicle is running, for example, blind spot monitoring or adaptive cruise control. Second, some systems transmit when in standby mode and not in use, making determination of actual usage difficult. Third, sensor calibration and positioning are extremely difficult to accomplish in the field. Fourth, many manufacturers are moving to vision-based systems for their automation technologies. As vision-based systems do not transmit light waves but instead read existing light waves, there is no way to detect whether they are installed or in use.

### 3.4.2 Leveraging Crash Databases

Crash studies present an alternative method for estimating ADAS usage rates. While crashes may not perfectly represent typical driving behavior, they can serve as a reasonable proxy. Crash investigations typically involve detailed examination of driver behavior and vehicle technologies, potentially yielding valuable data on ADAS usage.

This approach involves comparing ADAS activation rates in crashes to those in non-crash scenarios for equipped vehicles. If investigators can determine whether a driver was using an automation feature immediately prior to a crash, these rates could be compared against crashes involving equipped vehicles where the technology was not in use.

Several states have incorporated fields for ADAS usage in their crash report forms. The effectiveness of this data collection method remains uncertain, particularly regarding the accuracy of information gathered at crash scenes. Factors such as driver recall, system complexity, and investigator training may impact data quality.

### 3.4.3 Event Data Recorders

Electronic data recorders (EDRs) offer another potential avenue for estimating ADAS usage rates. Most new vehicles are voluntarily equipped with EDRs. For vehicles so equipped, they are required to record certain data at specified intervals in the seconds immediately prior to an activation event (*27*). Although current rules do not require EDRs to collect data on vehicle automation status, new rules could require the collection of on/off status of various automation features. Most Level 2 ADAS vehicles involve the integration of separate ADAS functions such as lane keeping, lane centering, adaptive cruise control, and collision avoidance.

Under the FAST Act of 2015, Congress permits the downloading of EDR data for research purposes provided any personally identifiable information and VIN are not disclosed (*28*). States could require crash investigations to download and record EDR data as part of police investigations into all crashes. Because manufacturers have nonstandard and often complex means to download EDR data, NHTSA could require manufacturers to install simple, uniform methods of data retrieval. A USB port installed under the dash of all new vehicles to seamlessly download EDR data is just one example. In the absence of comprehensive EDR data collection, a large sample of EDR data can be accessed through the Crash Investigative Sampling System (CISS) (*29*).

## 4    Discussion

The estimation of ADAS penetration rates presents significant challenges for transportation agencies and researchers. This study's methods and findings highlight both progress and substantial obstacles in estimating these rates.

### 4.1    Challenges in Current Estimation Methods

The HLDI approach, leveraging manufacturer-provided VIN data, represents the most sophisticated method for estimating ADAS penetration rates. This method overcomes many limitations inherent in relying solely on publicly available data. By obtaining VIN-level information directly from manufacturers, HLDI can more accurately determine which vehicles are equipped with optional ADAS features, a task otherwise extremely challenging.

The HLDI method is not without limitations. Reliance on voluntary manufacturer cooperation means the data may not be comprehensive across all makes and models. The proprietary nature of the data limits its availability to the broader research community. While HLDI publishes aggregate estimates, the underlying data is not accessible for independent analysis or for estimating penetration rates of ADAS features not included in HLDI's reports.

VIN decoding using NHTSA's vPIC database offers a more accessible alternative, but it too has significant limitations. The database only covers model years 2017 and newer, leaving a substantial portion of the vehicle fleet unaccounted for. Additionally, for optional ADAS features, this method can only determine if a feature was available, not if it was actually installed.

### 4.2 Challenges in Estimating Actual Usage

While estimating the penetration of installed ADAS features is crucial, understanding actual usage rates presents an even greater challenge. The limited studies available suggest activation rates vary widely between different ADAS features. This variability underscores the importance of distinguishing between equipped rates and actual usage rates when assessing the potential impact of ADAS on road safety and capacity.

The proposed approaches for estimating actual usage, such as naturalistic driving studies and enhanced event data recorder (EDR) data collection, offer promising avenues for addressing this knowledge gap. These methods also present significant challenges in terms of cost, privacy concerns, and the need for regulatory changes.

### 4.3 Implications for Transportation Planning and Policy

The current limitations in ADAS penetration rate estimation have significant implications for transportation planning and policy. Accurate estimates are crucial for calibrating traffic models, forecasting safety improvements, and making informed infrastructure investment decisions. The potential impacts of ADAS on road capacity (*1*) and safety (*2*, *3*) underscore the importance of reliable penetration rate data.

The lack of standardized, comprehensive data on ADAS penetration and usage also poses challenges for policymakers. Without accurate information on the prevalence and effectiveness of these technologies, it becomes difficult to develop evidence-based regulations or to assess the impact of existing policies aimed at promoting ADAS adoption.

### 4.4 Future Research Directions

Continued refinement of methods like those employed by HLDI, combined with efforts to make such data more widely available, could significantly improve the accuracy and accessibility of penetration rate estimates. Simultaneously, pursuing alternative data collection methods, such as enhanced EDR data or large-scale naturalistic driving studies, could provide crucial insights into actual ADAS usage patterns.

Standardization of ADAS feature definitions and data collection methods across manufacturers and regulatory agencies could also greatly facilitate more accurate and comparable penetration rate estimates. This standardization could potentially be achieved through collaborative efforts between industry stakeholders, researchers, and government agencies.

While significant progress has been made in estimating ADAS penetration rates, substantial challenges remain. Addressing these challenges will require continued innovation in data collection methods, increased collaboration between stakeholders, and potentially new regulatory frameworks to ensure the availability of comprehensive, accurate ADAS penetration and usage data.

## Acknowledgements

This work was sponsored by the Virginia Department of Transportation. The views and opinions expressed in this article are those of the author and do not necessarily reflect the official policies or positions of any agency of the Commonwealth of Virginia. The author thanks Amanda Hamm, Mena Lockwood, Angela Schneider, Patrick Harrison, Bridget Donaldson, and Michael Fontaine for providing useful input throughout the project.